\documentclass[a4paper, 11pt]{article}
\usepackage{amsfonts}
\usepackage{dsfont}
\usepackage{mathrsfs}
\usepackage{amssymb}
\usepackage{bbm}
\usepackage{epsfig}
\usepackage{amsmath}
\usepackage{float}
\usepackage{color}
\usepackage[english]{babel}
\usepackage{cite}
\usepackage{subfigure}
\usepackage{enumerate}

\usepackage{graphicx}

\def \defn{\triangleq}

\newcommand{\T}{^{\mbox{\tiny T}}}

\newtheorem{theorem}{Theorem}
\newtheorem{assumption}{Assumption}

\newtheorem{lemma}{Lemma}

\newtheorem{definition}{Definition}
\newtheorem{remark}{Remark}

\newenvironment{IEEEproof}[1][\bf Proof]%
{\smallskip\par\noindent\textit{#1: }}%
{\hspace*{\fill} \rule{6pt}{6pt}\smallskip}
\newenvironment{proof*}[1][Proof]%
{\smallskip\par\noindent\textbf{#1: }}%
{\smallskip}
\allowdisplaybreaks

\def\R{\mathbb{R}}

\newcommand{\eps}{\varepsilon}

  {\setlength{\arraycolsep}{0.5mm}\left\{ \; \begin{array}{#1}}%
      {\end{array} \right.}

\begin{document}
\title{Network Synchronization  with Nonlinear Dynamics  and Switching Interactions\thanks{This work has been supported in part by the Knut and Alice Wallenberg Foundation and the Swedish Research
Council.}}
\date{}

\author{Tao Yang\thanks{T. Yang is with the Pacific Northwest National Laboratory, 902 Battelle Boulevard, Richland, WA 99352 USA (e-mail: Tao.Yang@pnnl.gov).}, Ziyang Meng\thanks{Z. Meng is with the Institute for Information-Oriented Control, Technische Universit\"{a}t M\"{u}nchen, D-80290 Munich, Germany (e-mail: zmeng@lsr.ei.tum.de).}, Guodong Shi\thanks{G. Shi is with the College of Engineering and Computer Science, The Australian National University, Canberra ACT 0200, Australia (e-mail: guodong.shi@anu.edu.au).}, Yiguang Hong\thanks{Y. Hong is with the Key Laboratory of Systems and Control, Institute of Systems Science, Chinese Academy of Science, Beijing 100190, China (e-mail: yghong@iss.ac.cn).} and Karl Henrik Johansson\thanks{K. H. Johansson is with the ACCESS Linnaeus Centre, School of Electrical Engineering, Royal Institute of Technology, Stockholm 10044, Sweden (e-mail: kallej@kth.se).}
}

\maketitle

\begin{abstract}
This paper considers the synchronization problem for networks of coupled nonlinear dynamical systems under switching communication topologies. Two types of nonlinear agent dynamics are considered.
The first one is non-expansive dynamics (stable dynamics with a convex Lyapunov function $\varphi(\cdot)$) and the second one is dynamics that satisfies a global Lipschitz condition.
For the non-expansive case, we show that various forms of joint connectivity for communication graphs are
sufficient for networks to achieve global asymptotic $\varphi$-synchronization.
We also show that $\varphi$-synchronization leads to state synchronization provided that certain additional conditions are satisfied.
For the globally Lipschitz case, unlike the non-expansive case, joint connectivity alone is not sufficient for achieving synchronization. A sufficient condition for reaching global exponential synchronization is established in terms of the relationship between the global Lipschitz constant and the network parameters. We also extend the results to leader-follower networks. 
\end{abstract}


\section{Introduction} \label{sec-intro}
We consider the synchronization problem for a network of coupled nonlinear agents
with agent set $\mathcal{V}=\{1,2,\ldots, N\}$.
Their interactions (communications in the network) are described
by a time-varying directed graph $\mathcal{G}_{\sigma(t)}=(\mathcal{V}, \mathcal{E}_{\sigma(t)})$,
with $\sigma: [0,\infty) \rightarrow  \mathcal{P}$ as a piecewise constant signal, where $\mathcal{P}$ is a finite set of all possible graphs over $\mathcal{V}$.
The state of agent $i \in \mathcal{V}$ at time $t$ is denoted as $x_i(t) \in \R^n$ and
evolves according to
\begin{equation}\label{multi-agent-closed-leaderless}
\dot{x}_i = f(t,x_i) + \sum _{j \in \mathcal{N}_i(\sigma(t))} a_{ij}(t)(x_j-x_i),
\end{equation}
where $f(t,x_i): [0,\infty) \times {\R}^n  \rightarrow {\R}^n$ is piecewise continuous in $t$ and continuous in $x_i$
representing the uncoupled inherent agent dynamics,
$\mathcal{N}_i(\sigma(t))$ is the set of agent $i$'s neighbors at time $t$,
and $a_{ij}(t)>0$ is a piecewise continuous function marking the weight of edge $(j,i)$ at time $t$.

Systems of the form \eqref{multi-agent-closed-leaderless} have attracted considerable attention.
Most works focus on the case where the communication graph $\mathcal{G}_{\sigma(t)}$ is fixed, e.g.,\cite{wu-chua,wu-book,belykh-belykh-hasler,delellis-circuit,yu-chen-cao-tac2011,munz3,liu-cao-wu}. It is shown that
for the case where $f(t,x_i)$ satisfies a Lipschitz condition,
synchronization is achieved for a connected graph provided that the coupling strength is sufficiently large.
However, for the case where the communication graph is time-varying, the synchronization problem becomes much more challenging and existing literature mainly focuses on a few special cases when $f(t,x_i)$ is linear, e.g., the single-integrator case \cite{jadbabaie-lin-morse,moreau,lin07}, the double-integrator case \cite{ren-beard}, and the neutrally stable case \cite{scardovi-sepulchre,su-huang-tac}. Other studies assume some particular structures for the communication graph \cite{zhao-hill-liu,qin-gao-zheng,wen-duan-chen-yu}.
In particular, in \cite{zhao-hill-liu}, the authors focus on the case where the adjacency matrices associated with all communication graphs are simultaneously triangularizable. The authors of \cite{qin-gao-zheng} consider switching communication graphs that are weakly connected and balanced at all times. A more general case where the switching communication graph frequently has a directed spanning tree has been considered in \cite{wen-duan-chen-yu}. These special structures on the switching communication graph are rather restrictive compared to joint connectivity where the communication can be lost at any time.

This paper aims to investigate whether joint connectivity for switching communication graphs
can render synchronization for the nonlinear dynamics \eqref{multi-agent-closed-leaderless}.
We distinguish two classes depending on whether the nonlinear agent dynamics $f(t,x_i)$ is expansive or not.
For the non-expansive case, we focus on the case where the nonlinear agent dynamics is stable with a convex Lyapunov function $\varphi(\cdot)$. We show that various forms of joint connectivity for communication graphs are
sufficient for networks to achieve global asymptotic $\varphi$-synchronization, that is, the function $\varphi$ of the agent state converges to a common value.
We also show that $\varphi$-synchronization implies state synchronization provided that additional conditions are satisfied.
For the expansive case, we focus on when the nonlinear agent dynamics is globally Lipschitz, and
establish a sufficient condition for networks to achieve global exponential synchronization in terms of a relationship between the Lipschitz constant and the network parameters.

The remainder of this paper is organized as follows. Section~\ref{sec-prob} presents the problem definition and main results. Section~\ref{sec-proofs} provides technical proofs. In Section~\ref{sec-lf}, we extend the results to leader-follower networks. Finally, Section~\ref{sec-conclusion} concludes the paper. 

\section{Problem Definition and Main Results}\label{sec-prob}
\subsection{Problem Set-up}
Throughout the paper we make a standard dwell time assumption \cite{liberzon-morse} on  the switching signal $\sigma(t)$: there is a lower bound $\tau_D>0$ between two consecutive switching time instants of $\sigma(t)$. We also assume that
there are constants $0<a_{*} \leq a^{*}$ such that
$a_{*} \leq a_{ij}(t)\leq a^{*}$ for all $t \geq 0$.
We denote $x=[x\T_1,x\T_2,\ldots,x\T_N]\T \in \R^{nN}$ and
assume that the initial time is $t=t_0\geq 0$, and the initial state $x(t_0)=(x\T_1(t_0),\ldots,x\T_N(t_0))\T \in {\R}^{nN}$.
A digraph is {\it strongly connected} if it contains a directed path from every node to every other node.
The joint graph of $\mathcal{G}_{\sigma(t)}$ in the time interval $[t_1,t_2)$ with $t_1 < t_2 \leq \infty$ is denoted as $\mathcal{G}([t_1,t_2))=\cup _{t \in [t_1,t_2)} \mathcal{G}(t)=(\mathcal{V}, \cup_{t \in [t_1,t_2)} \mathcal{E}_{\sigma(t)})$.
For the communication graph, we introduce the following definition.

\begin{definition}\label{def-connected-leaderless}
(i). $\mathcal{G}_{\sigma(t)}$ is {\it uniformly jointly strongly connected} if there exists a constant $T>0$ such that $\mathcal{G}([t,t+T))$ is strongly connected for any $t \geq 0$.

(ii). Assume that $\mathcal{G}_{\sigma(t)}$ is undirected for all $t \geq 0$.
$\mathcal{G}_{\sigma(t)}$ is {\it infinitely jointly connected} if $\mathcal{G}([t,\infty))$ is connected for any $t \geq 0$.
\end{definition}

In this paper, we are interested in the following synchronization problems.

\begin{definition}\label{synch-def}
The multi-agent system \eqref{multi-agent-closed-leaderless} achieves global asymptotic $\varphi$-synchronization,
where $\varphi: {\R}^n \rightarrow \R$ is a continuously differentiable function, if
for any initial state $x(t_0)$, there exists a constant $d_\star(x(t_0))$, such that
$\lim_{t \rightarrow \infty} \varphi(x_i(t)) =d_\star(x(t_0))$ for any $i \in \mathcal{V}$ and any $t_0 \geq 0$.
\end{definition}

\begin{definition}\label{syn-asy}
(i) The multi-agent system \eqref{multi-agent-closed-leaderless} achieves global asymptotic synchronization if\\ 
$\lim_{t\to \infty}(x_i(t)-x_j(t))=0$ for any $i,j\in\mathcal{V}$, any $t_0 \geq 0$ and any $x(t_0) \in {\R}^{nN}$.

(ii) Multi-agent system \eqref{multi-agent-closed-leaderless} achieves global exponential synchronization if there exist $\gamma \geq 1$ and $\lambda>0$ such that
\begin{align}
\max_{\{i,j\}\in\mathcal{V}\times\mathcal{V}} \|x_i(t)-x_j(t)\|^2 \leq \gamma e^{-\lambda(t-t_0)} \max_{\{i,j\}\in\mathcal{V}\times\mathcal{V}} \|x_i(t_0)-x_j(t_0)\|^2, \, t\geq t_0,\label{global-expo-LL}
\end{align}
for any $t_0 \geq 0$ and any $x(t_0) \in {\R}^{nN}$.
\end{definition}

\begin{remark}
$\varphi$-synchronization is a type of output synchronization where the output of agent $i \in \mathcal{V}$ is chosen to be $\varphi(x_i)$. It is related to but different from $\chi$-synchronization \cite{cortes-auto08,hong-JSSC2010} since
$\varphi$ is a function of an individual agent state while $\chi$ is a function of all agent states.
\end{remark}

\subsection{Non-expansive Inherent Dynamics}\label{sec-ll}
In this section, we focus on when the nonlinear inherent agent dynamics is non-expansive as indicated by the following assumption.

\begin{assumption}\label{ass-lyap}
$\varphi: {\R}^n \rightarrow \R$ is a continuously differentiable positive definite convex function
satisfying
\begin{itemize}
\item[(i).] $\lim_{\|\eta\| \rightarrow \infty}\varphi(\eta)=\infty$;
\item[(ii).] $\langle \nabla \varphi(\eta), f(t,\eta) \rangle \leq 0$ for any $\eta \in {\R}^n$ and any $t \geq 0$.
\end{itemize}
\end{assumption}

The following lemma shows how Assumption  \ref{ass-lyap} enforces non-expansive dynamics.
\begin{lemma}\label{lem-invariant-leaderless}
Let Assumption~\ref{ass-lyap} hold. Along the multi-agent dynamics \eqref{multi-agent-closed-leaderless},
$\max_{i \in \mathcal{V}} \varphi(x_i(t))$ is non-increasing for all $t \geq 0$.
\end{lemma}

We now state main results for the non-expansive case.
\begin{theorem}\label{thm1-leaderless}
Let Assumption~\ref{ass-lyap} hold.
The multi-agent system \eqref{multi-agent-closed-leaderless} achieves
global asymptotic $\varphi$-synchronization if $\mathcal{G}_{\sigma(t)}$ is uniformly jointly strongly connected.
\end{theorem}

\begin{theorem}\label{thm2-leaderless}
Let Assumption~\ref{ass-lyap} hold.
Assume that $\mathcal{G}_{\sigma(t)}$ is undirected for all $t \geq t_0$.
The multi-agent system \eqref{multi-agent-closed-leaderless} achieves global asymptotic $\varphi$-synchronization if $\mathcal{G}_{\sigma(t)}$ is infinitely jointly connected.
\end{theorem}

\begin{remark}\label{remark-ass-stable}
For the linear time-varying case $f(t,x)=A(t) x$, if there exists a matrix $P=P\T>0$ such that
\begin{equation}\label{LTV-common}
P A(t)+A\T(t)P \leq 0, \quad \forall t \geq 0,
\end{equation}
then $\varphi(x)=x\T P x$ for $x\in {\R}^n$ satisfies Assumption~\ref{ass-lyap}.
For the linear time-invariant case $f(t,x)=A x$, the condition \eqref{LTV-common} is
equivalent to that the matrix $A$ is neutrally stable \cite{su-huang-tac}.
\end{remark}


\subsection{$\varphi$-synchronization vs. State Synchronization}\label{discussion-phi-state}
The following result establishes conditions under which $\varphi$-synchronization may imply state synchronization.

\begin{theorem}\label{prop-guodong}
Let $\mathcal{G}_{\sigma(t)}\equiv \mathcal{G}$ with $\mathcal{G}$ being a fixed, strongly connected digraph under which the multi-agent system \eqref{multi-agent-closed-leaderless} achieves global asymptotic $\varphi$-synchronization
for some positive definite function $\varphi: {\R}^n \rightarrow \R$.
Let Assumption \ref{ass-lyap} hold.
Moreover, assume that
\begin{enumerate}[(i).]
\item $f(t,\eta)$ is bounded for any $t \geq 0$ and any $\eta \in {\R}^n$.
\item $c_1 \|\eta\|^2 \leq \varphi(\eta) \leq c_2 \|\eta\|^2$ for some $0<c_1\leq c_2$; and
\item $\varphi(\cdot)$ is strongly convex.
\end{enumerate}
%
%
Then the multi-agent system \eqref{multi-agent-closed-leaderless} achieves global asymptotic synchronization.
\end{theorem}

\subsection{Lipschitz Inherent Dynamics}\label{sec-unstable}
We consider also the case when the nonlinear inherent agent dynamics is possibly expansive.
We focus on when the dynamics satisfies
the following global Lipschitz condition.

\begin{assumption}\label{ass-phi-state-1}
There exists a constant $L>0$ such that
\begin{equation}\label{ass-unstable-nonlinear}
\|f(t,\eta)-f(t,\zeta)\| \leq L \|\eta-\zeta\|, \quad \forall \eta, \zeta \in {\R}^n, \, \forall t \geq 0.
\end{equation}
\end{assumption}

Our  main result for this case is given below.

\begin{theorem}\label{unstable-thm1}
Let Assumption~\ref{ass-phi-state-1} hold.
Assume that $\mathcal{G}_{\sigma(t)}$ is uniformly jointly strongly connected.
Global exponential synchronization is achieved for the
multi-agent system \eqref{multi-agent-closed-leaderless}
if $L< {\rho_*}/{2}$, where $\rho_*$ is a constant 
depending on the network parameters.
\end{theorem}

\begin{remark}\label{remark-unstable-ass}
Assumption~\ref{ass-phi-state-1} and its variants have been considered in the literature for fixed communication
graphs, e.g.,\cite{wu-chua,wu-book,belykh-belykh-hasler,delellis-circuit,liu-cao-wu,yu-chen-cao-tac2011,munz3}.
Compared with the existing literature, we here study a more challenging case, where the communication graphs are time-varying.
Unlike the fixed case where the global Lipschitz condition is sufficient to guarantee synchronization,
Theorem~\ref{unstable-thm1} established a sufficient synchronization condition related to the Lipschitz constant and the network parameters.
\end{remark}

\section{Proofs of the Main Results}\label{sec-proofs}
In this section, we provide proofs of the main results.
\subsection{Proof of Lemma~\ref{lem-invariant-leaderless}}\label{proof-lem-invariant-leaderless}
Recall that the upper Dini derivative of a function $h(t): (a,b)\to {\R}$ at $t$ is defined as $D^+h(t)=\limsup_{s\to 0^+} \frac{h(t+s)-h(t)}{s}$. The following lemma from \cite{dan,lin07} is useful for the proof.
\begin{lemma}\label{lem1}
Let $V_i(t,x): {\R}\times {\R}^n \to \R \;(i=1,\dots,N)$ be
continuously differentiable and $V(t,x)=\max_{i=1,\dots,N}V_i(t,x)$. If $
\mathcal{I}(t)=\{i\in \{1,2,\dots,N\}\,:\,V(t,x(t))=V_i(t,x(t))\}$
is the set of indices where the maximum is reached at $t$, then $D^+V(t,x(t))=\max_{i\in\mathcal{I}(t)}\dot{V}_i(t,x(t))$.
\end{lemma}
Denote $\mathcal{I}(t)=\{i \in \mathcal{V}: \max_{i \in \mathcal{V}} \varphi(x_i(t))=\varphi(x_i(t))\}$.
We first note that the convexity property of $\varphi(\cdot)$ implies that \cite[pp.69]{boyd-vandenberghe}
\begin{equation}\label{convexity-eq}
\langle \nabla \varphi(\eta), \zeta-\eta \rangle \leq \varphi(\zeta)-\varphi(\eta), \quad \forall \eta,\zeta \in {\R}^n.
\end{equation}
It then follows from Lemma \ref{lem1}, Assumption~\ref{ass-lyap}(ii) and \eqref{convexity-eq} that
\begin{align*}
D^+ \max_{i \in \mathcal{V}} \varphi(x_i(t))&= \max_{i \in \mathcal{I}(t)} \big \langle \nabla \varphi(x_i), f(t,x_i)+\sum _{j \in \mathcal{N}_i(\sigma(t))} a_{ij}(t)(x_j-x_i) \big \rangle \\
& \leq \max_{i \in \mathcal{I}(t)} \sum _{j \in \mathcal{N}_i(\sigma(t))}a_{ij}(t)(\varphi(x_j)-\varphi(x_i)) \leq 0,
\end{align*}
where the last inequality follows from $\varphi(x_j) \leq \varphi(x_i)$. This proves the lemma.

\subsection{Proof of Theorem~\ref{thm1-leaderless}}\label{proof-thm1-leaderless}
It follows from Lemma~\ref{lem-invariant-leaderless} that for any initial state $x(t_0) \in {\R}^{nN}$,
there exists a constant $d_*=d_\star(x(t_0)) \geq 0$, such that $\lim_{t \rightarrow \infty} \max_{i \in \mathcal{V}} \varphi(x_i) =d_\star$. We shall show that $d_\star$ is the required constant in Definition \ref{synch-def} of $\varphi$-synchronization.

We first note that by Lemma~\ref{lem-invariant-leaderless} that for all $i \in \mathcal{V}$, there exist constants $0 \leq \alpha_i \leq \beta_i \leq d_\star$,
such that
\[
\liminf_{t \rightarrow \infty} \varphi(x_i(t))=\alpha_i, \quad \limsup_{t \rightarrow \infty} \varphi(x_i(t))=\beta_i.
\]
Also note that it follows from $\lim_{t \rightarrow \infty}  \max_{i \in \mathcal{V}}\varphi(x_i(t))=d_\star$ that for any $\eps>0$, there exists $T_1(\eps)>0$ such that
\begin{equation}\label{bound-T1}
\varphi(x_i(t)) \in [0, d_\star+\eps], \quad \forall i \in \mathcal{V}, \, \forall t \geq T_1(\eps).
\end{equation}

The proof of Theorem~\ref{thm1-leaderless} is based on a contradiction argument and
relies on the following lemma.
\begin{lemma}\label{lem1-thm1}
Let Assumption~\ref{ass-lyap} hold. Assume that $\mathcal{G}_{\sigma(t)}$ is uniformly jointly strongly connected.
If there exists an agent $k_0 \in \mathcal{V}$ such that
$0 \leq \alpha_{k_0}< d_\star$, then there exists $0<\bar{\rho}<1$ and $\bar{t}$ such that for all $i \in \mathcal{V}$,
$\varphi(x_i(\bar{t}+(N-1)T_0)) \leq \bar{\rho}M_0+(1-\bar{\rho})(d_\star+\varepsilon)$, where
\begin{equation}\label{defn-T0}
T_0 \defn T+2\tau_D,
\end{equation}
with $T$ given in Definition \ref{def-connected-leaderless}(i) and $\tau_D$ is the dwell time.
\end{lemma}
\begin{IEEEproof}
Let us first define $M_0 \defn \frac{\alpha_{k_0}+\beta_{k_0}}{2}< d_\star$.
Then there exists an infinite time sequence $t_0<\tilde{t}_1<\ldots<\tilde{t}_k < \ldots$
with $\lim_{k \rightarrow \infty}\tilde{t}_k=\infty$ such that $\varphi(x(\tilde{t}_k))=M_0$ for all $k=1,2,\ldots$.
We then pick up one $\tilde{t}_k$, $k=1,2,\ldots$ such that it is greater than or equal to $T_1(\eps)$
and denote it as $\tilde{t}_{k_0}$.

We now prove the lemma by estimating an upper bound of the scalar function $\varphi(x_i)$ agent by agent.
The proof is based on a generalization of the method proposed in the proof of \cite[Lemma~4.3]{shi-johansson-hong}
but with substantial differences on the agent dynamics and Lyapunov function. Moreover, the convexity of $\varphi(\cdot)$ plays an important role.

\noindent Step 1. Focus on agent $k_0$.
By using Assumption~\ref{ass-lyap}(ii), \eqref{convexity-eq}, and \eqref{bound-T1}, we obtain that for all $t \geq \tilde{t}_{k_0}$,
\begin{align}
\frac{d}{dt} \varphi(x_{k_0}(t))&=\big \langle \nabla \varphi(x_{k_0}), f(t,x_{k_0})+  \sum_{j \in \mathcal{N}_{k_0}(\sigma(t))}a_{k_0 j}(t)(x_j-x_{k_0}) \big \rangle  \nonumber \\
& \leq \sum _{j \in \mathcal{N}_{k_0}(\sigma(t))} a_{k_0 j}(t)  \left(\varphi(x_j)-\varphi(x_{k_0})\right) \nonumber\\
& \leq a^* (N-1) (d_\star+\eps -\varphi(x_{k_0})).\label{leaderless-k1}
\end{align}
It then follows that for all $t \geq \tilde{t}_{k_0}$,
\begin{equation}
\varphi(x_{k_0}(t)) \leq e^{-\lambda_1(t-\tilde{t}_{k_0})}  \varphi(x_{k_0}(\tilde{t}_{k_0}))+\left(1-e^{-\lambda_1(t-\tilde{t}_{k_0})}\right)(d_\star+\eps),\label{dis-k0}
\end{equation}
where $\lambda_1=a^* (N-1)$.

\vspace{2mm}

\noindent Step 2. Consider agent $k_1 \neq k_0$ such that $(k_0,k_1) \in \mathcal{E}_{\sigma(t)}$ for
$t \in [\tilde{t}_{k_0},\tilde{t}_{k_0}+T_0)$. The existence of such an agent can be shown as follows.
Since $\mathcal{G}_{\sigma(t)}$ is uniformly jointly strongly connected, it is not hard to see
that there exists an agent $k_1 \neq k_0 \in \mathcal{V}$ and $t_1 \geq \tilde{t}_{k_0}$ such that
$(k_0,k_1) \in \mathcal{E}_{\sigma(t)}$ for $t \in [t_1,t_1+\tau_D) \subseteq [\tilde{t}_{k_0},\tilde{t}_{k_0}+T_0)$.

From \eqref{dis-k0}, we obtain for all $t \in [\tilde{t}_{k_0},\tilde{t}_{k_0}+(N-1)T_0]$,
\begin{equation}\label{inner-point}
\varphi(x_{k_0}(t))  \leq \kappa_0 \defn \rho M_0+(1-\rho)(d_\star+\eps),
\end{equation}
where $\rho=e^{-\lambda_1(N-1)T_0}=e^{-a^* (N-1)^2 T_0}$.

We next estimate $\varphi(x_{k_1}(t))$ by considering two different cases.

Case I: $\varphi(x_{k_1}(t))>\varphi(x_{k_0}(t))$ for all $t \in [t_1,t_1+\tau_D)$.

By using Assumption~\ref{ass-lyap}(ii), \eqref{convexity-eq}, \eqref{bound-T1}, and \eqref{inner-point}, we obtain for all $t \in [t_1,t_1+\tau_D)$,
\begin{align*}
\frac{d}{dt} \varphi(x_{k_1}(t)) & \leq \sum _{j \in \mathcal{N}_{k_1}(\sigma(t)) \setminus \{k_0\}}
a_{k_1 j}(t)\left(\varphi(x_{k_j})-\varphi(x_{k_1})\right) +
a_{k_1 k_0}(t)(\varphi(x_{k_0})-\varphi(x_{k_1}))\\
& \leq a^*(N-2) (d_\star+\eps -\varphi(x_{k_1}))+a_*\left(\kappa_0-\varphi(x_{k_1})\right).
\end{align*}

From the preceding relation, we obtain for $t \in [t_1,t_1+\tau_D)$,
\begin{align*}
\varphi(x_{k_1}(t)) \leq e^{-\lambda_2 (t-t_1)}  \varphi(x_{k_1}(t_1))+\frac{\left[a^*(N-2)(d_\star+\eps)+a_*\kappa_0\right](1-e^{-\lambda_2(t-t_1)})}{\lambda_2},
\end{align*}
where $\lambda_2=a^*(N-2)+a_*$.
Therefore, we have
\begin{equation}\label{inner-point-2}
\varphi(x_{k_1}(t_1+\tau_D)) \leq \kappa_1  \defn \mu(d_\star+\eps)+(1-\mu)\kappa_0,
\end{equation}
where
\begin{equation}\label{def-mu}
\mu=\frac{\lambda_2-a_*(1-e^{-\lambda_2\tau_D})}{\lambda_2}.
\end{equation}
By applying the same analysis as we obtained \eqref{dis-k0} to the agent $k_1$,
we obtain for all $t \geq t_1+\tau_D$,
\begin{equation}\label{inner-point-i1}
\varphi(x_{k_1}(t))  \leq e^{-\lambda_1(t-(t_1+\tau_D))}\kappa_1+\Big[1- e^{-\lambda_1(t-(t_1+\tau_D))}\Big](d_\star+\eps).
\end{equation}
By combining the inequalities \eqref{inner-point}, \eqref{inner-point-2} and \eqref{inner-point-i1}, we obtain for all $t \in [t_1+\tau_D, \tilde{t}_{k_0}+(N-1)T_0]$,
\begin{equation}\label{inner-key}
\varphi(x_{k_1}(t))  \leq \varphi_1 M_0+(1-\varphi_1)(d_\star+\eps),
\end{equation}
where $\varphi_1=(1-\mu)\rho^2 $.

Case II: There exists a time instant $\bar{t}_1 \in [t_1,t_1+\tau_D)$ such that
\begin{equation}\label{k1-case2}
\varphi(x_{k_1}(\bar{t}_1)) \leq \varphi(x_{k_0}(\bar{t}_1))\leq \kappa_0.
\end{equation}
By applying the similar analysis as we obtained \eqref{leaderless-k1}
to the agent $k_1$, we obtain for all $t \geq \tilde{t}_{k_0}$,
\[
\frac{d}{dt} \varphi(x_{k_1}(t))  \leq  a^* (N-1) (d_\star+\eps -\varphi(x_{k_1}(t))).
\]
This leads to
\[
\varphi(x_{k_1}(t))  \leq e^{-\lambda_1(t-\bar{t}_1)}  \varphi(x_{k_1}(\bar{t}_1))+(1-e^{-\lambda_1(t-\bar{t}_1)})(d_\star+\eps).
\]
By combining the preceding relation, \eqref{inner-point}, and \eqref{k1-case2}, and using $0<\varphi_1=(1-\mu)\rho^2<\rho^2$ which follows from $0<\mu<1$, we obtain for all  $t \in [t_1+\tau_D, \tilde{t}_{k_0}+(N-1)T_0]$,
\begin{equation*}
\varphi(x_{k_1}(t)) \leq \rho^2 M_0+(1-\rho^2)(d_\star+\eps) < \varphi_1 M_0+(1-\varphi_1)(d_\star+\eps). \label{inner-key1}
\end{equation*}
From the preceding relation and \eqref{inner-key}, it follows that for both cases, we have
for all  $t \in [t_1+\tau_D, \tilde{t}_{k_0}+(N-1)T_0]$,
\begin{equation*}\label{inner-key2}
\varphi(x_{k_1}(t)) \leq \varphi_1 M_0+(1-\varphi_1)(d_\star+\eps).
\end{equation*}
From the preceding relation, \eqref{inner-point} and $0<\varphi_1<\rho<1$,
it follows that for all $t \in [t_1+\tau_D, \tilde{t}_{k_0}+(N-1)T_0]$,
\begin{equation}\label{inner-key-01}
\varphi(x_j(t))  \leq \varphi_1 M_0+(1-\varphi_1)(d_\star+\eps), \quad j \in \{k_0,k_1\}.
\end{equation}

\vspace{2mm}

\noindent Step 3. Consider agent $k_2 \notin \{k_0,k_1\}$ such that there exists an edge from the set $\{k_0,k_1\}$ to the agent $k_2$ in $\mathcal{E}_{\sigma(t)}$ for $t \in [t_2,t_2+\tau_D) \subseteq [\tilde{t}_{k_0}+T_0,\tilde{t}_{k_0}+2T_0)$.
The existence of such an agent $k_2$ and $t_2$ follows similarly from the argument in Step 2.

Similarly, we can bound $\varphi(x_{k_2}(t))$ by considering two different cases and obtain that
for all $t \in [t_2+\tau_D, \tilde{t}_{k_0}+(N-1)T_0]$,
\begin{equation}\label{inner-key-2}
\varphi(x_{k_2}(t)) \leq \varphi_2 M_0+(1-\varphi_2)(d_\star+\eps),
\end{equation}
where $\varphi_2=((1-\mu)\rho^2)^2$.

By combining \eqref{inner-key-01} and \eqref{inner-key-2}, and using $0<\varphi_2<\varphi_1<1$,
we obtain that for all $t \in [t_2+\tau_D, \tilde{t}_{k_0}+(N-1)T_0]$,
\begin{equation*}\label{inner-key-012}
\varphi(x_j(t))\leq \varphi_2 M_0+(1-\varphi_2)(d_\star+\eps), \quad j \in \{k_0,k_1,k_2\}.
\end{equation*}

\vspace{2mm}

\noindent Step 4. By repeating the above process on
time intervals $[\tilde{t}_{k_0}+2T_0,\tilde{t}_{k_0}+3T_0), \; \ldots,\; [\tilde{t}_{k_0}+(N-2)T_0,\tilde{t}_{k_0}+(N-1)T_0)$,
we eventually obtain that for all $i \in \mathcal{V}$,
\begin{equation*}\label{inner-key-N}
\varphi(x_i(\tilde{t}_{k_0}+(N-1)T_0)) \leq \varphi_{N-1} M_0+(1-\varphi_{N-1})(d_\star+\eps).
\end{equation*}
where $\varphi_{N-1}=((1-\mu)\rho^2)^{N-1}$.
The result of the lemma then follows by choosing $\bar{\rho}=\varphi_{N-1}$ and $\bar{t}=\tilde{t}_{k_0}$.
\end{IEEEproof}

We are now ready to prove Theorem~\ref{thm1-leaderless} by contradiction.
Suppose that there exists an agent $k_0 \in \mathcal{V}$ such that
$0 \leq \alpha_{k_0}< d_\star$. It then follows from Lemma~\ref{lem1-thm1} that $\varphi(x_i(\bar{t}+(N-1)T_0))<d_\star$
for all $i \in \mathcal{V}$, provided that $\eps<\frac{\bar{\rho}(d_\star-M_0)}{1-\bar{\rho}}$.
This contradicts the fact that $\lim_{t \rightarrow \infty} \max_{i \in \mathcal{V}} \varphi(x_i) =d_\star$.
Thus, there does not exist an agent $k_0 \in \mathcal{V}$ such that $0 \leq \alpha_{k_0}<d_\star$. Hence, $\lim_{t \rightarrow \infty} \varphi(x_i(t)) =d_\star$ for all $i \in \mathcal{V}$.

\subsection{Proof of Theorem \ref{thm2-leaderless}}\label{proof-thm2-leaderless}
The proof relies on the following lemma.
\begin{lemma}\label{lem1-thm2}
Let Assumption~\ref{ass-lyap} hold. Assume that $\mathcal{G}_{\sigma(t)}$ is infinitely jointly connected.
If there exists an agent $k_0 \in \mathcal{V}$ such that
$0 \leq \alpha_{k_0}< d_\star$, then there exist $0<\tilde{\rho}<1$ and $\tilde{t}$ such that
\[
\varphi(x_i(\tilde{t}+\tau_D)) \leq \tilde{\rho}M_0+(1-\tilde{\rho})(d_\star+\varepsilon), \quad \, \forall i \in \mathcal{V}.
\]
\end{lemma}

\begin{IEEEproof}
The proof of Lemma~\ref{lem1-thm2} is similar to that of Lemma~\ref{lem1-thm1} and based on estimating
an upper bound for the scalar quantity $\varphi(x_i)$ agent by agent.
However, since $\mathcal{G}_{\sigma(t)}$ is infinitely jointly connected,
the method that we get the order of the agents based on the intervals induced by the uniform bound $T$ cannot be used here.
We can however apply the strategy in \cite{shi-johansson-hong} for the analysis as shown below: 

\noindent Step 1. In this step, we focus on agent $k_0$. Since $\mathcal{G}_{\sigma(t)}$ is infinitely jointly connected,
we can define
\begin{equation*}\label{IJC-def-t1}
\hat{t}_1 \defn \inf_{t \in [\tilde{t}_{k_0},\infty)} \{\exists \, i \in \mathcal{V} \mid (k_0,i) \in \mathcal{E}_{\sigma(t)}\},
\end{equation*}
and the set
\begin{equation*}\label{IJC-def-V1}
\mathcal{V}_1 \defn \{ i \in \mathcal{V} \mid (k_0,i) \in \mathcal{E}_{\sigma(\hat{t}_1)}\}.
\end{equation*}
For $\tilde{t}_{k_0} \leq t < \hat{t}_1$, agent $k_0$ has no neighbor, it follows from Assumption~\ref{ass-lyap}(ii) that
$\frac{d}{dt}\varphi(x_{k_0}(t)) =\langle \nabla \varphi(x_{k_0}), f(t,x_{k_0})\rangle \leq 0$.
Thus, $\varphi(x_{k_0}(t)) \leq \varphi(x_{k_0}(\tilde{t}_{k_0}))$ for $\tilde{t}_{k_0} \leq t < \hat{t}_1$.
By applying a similar analysis as we obtained \eqref{inner-point}, we have for all $t \in [\hat{t}_1,\hat{t}_1+\tau_D)$,
\begin{equation*}\label{inner-point-undirected}
\varphi(x_{k_0}(t)) \leq \hat{\kappa}_0 \defn \hat{\rho} M_0+(1-\hat{\rho})(d_\star+\eps),
\end{equation*}
where $\hat{\rho}=e^{-a^* (N-1)\tau_D}$.

\vspace{2mm}

\noindent Step 2. In this step, we focus on all $k_1 \in \mathcal{V}_1$.
We then estimate $\varphi(x_{k_1}(t))$ for all $k_1 \in \mathcal{V}_1$ by considering two different cases, i.e.,
Case I: If $\varphi(x_{k_1}(t))>\varphi(x_{k_0}(t))$ for all $t \in [\hat{t}_1,\hat{t}_1+\tau_D)$, and
Case II: If there exists a time instant $\bar{t}_1 \in [\hat{t}_1,\hat{t}_1+\tau_D)$ such that
$\varphi(x_{k_1}(\bar{t}_1)) \leq \varphi(x_{k_0}(\bar{t}_1))\leq \hat{\kappa}_0$.
By using the similar argument as the two-case analysis for agent $k_1$ in the proof of Theorem~\ref{thm1-leaderless},
we eventually obtain for all $j \in k_0 \cup \mathcal{V}_1$,
\begin{equation}\label{inner-point-3-undirected}
\varphi(x_j(\hat{t}_1+\tau_D))  < \hat{\varphi}_1M_0+(1-\hat{\varphi}_1)(d_\star+\eps),
\end{equation}
where $\hat{\varphi}_1=(1-\mu){\hat{\rho}}^2$ with $\mu$ given by \eqref{def-mu}.

\vspace{2mm}

\noindent Step 3. We then view the set $\{k_0\} \cup \mathcal{V}_1$ as a subsystem. Define $\hat{t}_2$
as the first time when there is an edge between this subsystem and the remaining agents and $\mathcal{V}_2$ accordingly.
By using the similar analysis for agent $k_1$ in Step 2, we can estimate the upper bound for all the agent in the set
$\{k_0\} \cup \mathcal{V}_1 \cup \mathcal{V}_2$,

\vspace{2mm}

\noindent Step 4. Since $\mathcal{G}_{\sigma(t)}$ is infinitely jointly connected, we can continue the above process until
$\mathcal{V}=\{k_0\} \cup \mathcal{V}_1 \cup \cdots \cup \mathcal{V}_\ell$ for some $\ell \leq N-1$.
Eventually, we have
\begin{equation*}\label{inner-key-N-undirected-key}
\varphi(x_{i}(\hat{t}_\ell+\tau_D)) <  \hat{\varphi}_{N-1} M_0+(1-\hat{\varphi}_{N-1})(d_\star+\eps), \quad \forall i \in \mathcal{V}.
\end{equation*}
The result of lemma then follows by choosing $\tilde{t}=\hat{t}_{\ell}$ and $\tilde{\rho}=\hat{\varphi}_{N-1}$.
\end{IEEEproof}

The remaining proof of Theorem~\ref{thm2-leaderless} follows from a contradiction argument
and Lemma~\ref{lem1-thm2} in the same way as the proof of Theorem~\ref{thm1-leaderless}.

\subsection{Proof of Theorem~\ref{prop-guodong}}\label{proof-prop-guodong}
If the multi-agent system \eqref{multi-agent-closed-leaderless}
reaches  asymptotic $\varphi$-synchronization, i.e.,
$\lim_{t \rightarrow \infty} \varphi(x_i(t)) =d_\star$  for all $i \in \mathcal{V}$, then for any $\epsilon >0$, there exists a $T_\epsilon>0$ such that
\begin{equation}\label{bound-phi}
d_\star -\epsilon\leq \varphi(x_i(t)) \leq d_\star +\epsilon, \quad \forall i \in \mathcal{V}, \quad \forall t \geq T_\epsilon.
\end{equation}

If $d_\star=0$ the desired conclusion holds trivially, i.e., $\lim_{t \rightarrow \infty}x_i(t)=0$ for all $i \in \mathcal{V}$
due to the positive definiteness of $\varphi(\cdot)$.
In the remainder of the proof we assume $d_\star >0$. We shall prove the result by contradiction. Suppose that state synchronization is not achieved, then there exist two agents $i_0,j_0 \in\mathcal{V}$ such that $\limsup_{t\to \infty} \|x_{i_0}(t)-x_{j_0}(t)\|>0$.
In other words, there exist an infinite time sequence $t_1<\dots<t_k<\dots$ with $\lim_{k\to \infty} t_k =\infty$, and a constant $\delta>0$ such that $\|x_{i_0}(t_k)-x_{j_0}(t_k)\|=2^{N-1}\sqrt{\delta/c_1}$ for all $k\geq 1$, where $c_1$ is given in condition (ii) of Theorem~\ref{prop-guodong}. We divide the following analysis into three steps.

\vspace{2mm}

\noindent Step 1. In this step, we prove the following crucial claim.

\vspace{1mm}

\noindent{\it Claim.} For any $t_k$, there are two agents $i_\ast, j_\ast \in\mathcal{V}$ with $(i_\ast,j_\ast)\in \mathcal{E}$ such that $\|x_{i_\ast}(t_k)-x_{j_\ast}(t_k)\|\geq \sqrt{\delta/c_1}$.

We establish this claim via a recursive analysis. If either $(i_0,j_0)\in  \mathcal{E}$ or  $(j_0,i_0)\in  \mathcal{E}$ then the result follows trivially. Otherwise we pick up another agent $k_0$ satisfying that there is an edge between $k_0$ and $\{i_0,j_0\}$. This $k_0$ always exists since $\mathcal{G}$ is strongly connected. Then either $\|x_{k_0}(t_k)-x_{i_0}(t_k)\|\geq 2^{N-2}\sqrt{\delta/c_1}$ or $\|x_{k_0}(t_k)-x_{j_0}(t_k)\|\geq 2^{N-2}\sqrt{\delta/c_1}$ must hold. Thus, we have again either established the claim, or we can continue to select another agent different from $i_0$, $j_0$, and $k_0$ and repeat the argument. Since we have a finite number of agents, the desired claim holds.

Furthermore, since there is a finite number of agent pairs, without loss of generality, we assume that the given agent pair $i_\ast,j_\ast$ does not vary for different $t_k$ (otherwise we can always select an infinite subsequence of $t_k$ for the following discussions).

\vspace{2mm}

\noindent Step 2. In this step, we establish a lower bound of $\|x_{i_\ast}(t)-x_{j_\ast}(t)\|^2$ for a small time interval after a particular $t_k$ satisfying $t_k> T_\epsilon$.
From \eqref{bound-phi} and $\lim_{\|x\| \rightarrow \infty}\varphi(x)=\infty$ given in Assumption~\ref{ass-lyap}(i),
we see that $x_i(t)$ and $\nabla \varphi(x_i(t)-x_j(t))$ are bounded for all $i, j\in\mathcal{V}$ and for all $t\geq T_\epsilon$.
It then follows from condition (i) of Theorem~\ref{prop-guodong} and \eqref{multi-agent-closed-leaderless} that
\begin{align*}
\Big|\frac{d}{dt}\varphi(x_{i_\ast}-x_{j_\ast})\Big| & \leq \Big| \big \langle \nabla \varphi(x_{i_\ast}-x_{j_\ast}), f(t,x_{i_\ast})-f(t,x_{j_\ast})\rangle\Big| \\
 &\hspace{0.4cm} +\Big| \big \langle \nabla \varphi(x_{i_\ast}-x_{j_\ast}), \sum _{k_1 \in \mathcal{N}_{i_\ast}(\sigma(t))} a_{i_\ast k_1}(t)(x_{k_1}-x_{i_\ast})\rangle\Big| \\
&\hspace{0.4cm}+\Big| \big \langle \nabla \varphi(x_{i_\ast}-x_{j_\ast}), \sum _{k_2 \in \mathcal{N}_{j_\ast}(\sigma(t))} a_{j_\ast k_2}(t)(x_{k_2}-x_{j_\ast})\rangle\Big| \leq L_\ast
\end{align*}
for all $t \geq t_k$ and some $L_\ast >0$.
Without loss of generality we assume that $\epsilon \leq 1$. Then $L_\ast$ will be independent of $\epsilon$.

By plugging in the fact that $\|x_{i_\ast}(t_k)-x_{j_\ast}(t_k)\| \geq \sqrt{\delta/c_1} $ and using the condition (ii) of Theorem~\ref{prop-guodong}, we obtain that
\begin{equation}\label{guodong1}
\|x_{i_\ast}(t)-x_{j_\ast}(t)\|^2  \geq \frac{\delta}{2c_2}, \quad \, t\in \left[t_k,t_k+ \frac{\delta}{2 L_\ast}\right].
\end{equation}

\vspace{2mm}

\noindent Step 3. We first note that the strong convexity of $\varphi(\cdot)$ implies that \cite[pp.459]{boyd-vandenberghe}
there exists an $m>0$ such that
\begin{equation}\label{strongly-convexity-eq}
\langle \nabla \varphi(\eta), \zeta-\eta \rangle \leq \varphi(\zeta)-\varphi(\eta)-\frac{m}{2}\|\eta-\zeta\|^2, \, \forall \eta,\zeta \in {\R}^n.
\end{equation}
By using Assumption~\ref{ass-lyap}(ii) and \eqref{strongly-convexity-eq}, we obtain for $t\geq T_\epsilon$,
\begin{align}\label{guodong2}
\frac{d}{dt}\varphi(x_{j_\ast}(t))
& \leq a_{j_\ast i_\ast }(t) \Big(\varphi(x_{i_\ast})-\varphi(x_{j_\ast})-\frac{m}{2}\|x_{i_\ast}-x_{j_\ast}\|^2\Big) \nonumber \\
& \hspace{0.4cm}+\sum _{k \in \mathcal{N}_{j_\ast}(\sigma(t))\setminus \{i_\ast\}} a_{j_\ast k}(t) \Big(\varphi(x_k)-\varphi(x_{j_\ast})\Big)\nonumber\\
& \leq  a_{j_\ast i_\ast }(t)  \Big|\varphi(x_{i_\ast})-\varphi(x_{j_\ast})\Big|
-\frac{m}{2}a_{j_\ast i_\ast }(t)\|x_{i_\ast}(t)-x_{j_\ast}(t)\|^2 \nonumber \\
& \hspace{0.4cm}+\sum _{k \in \mathcal{N}_{j_\ast}(\sigma(t))\setminus \{i_\ast\}} a_{j_\ast k}(t)\Big|\varphi(x_k)  -\varphi(x_{j_\ast})\Big|,
\end{align}
By using \eqref{bound-phi}, \eqref{guodong1}, \eqref{guodong2}, and condition (ii) of Theorem~\ref{prop-guodong},
we obtain that for $t\in \left[t_k,t_k+ \frac{\delta}{2 L_\ast}\right]$,
\begin{equation*}
\frac{d}{dt}\varphi(x_{j_\ast}(t)) \leq  2(N-1)a^\ast \epsilon -\frac{a_\ast m \delta}{4c_2}, 
\end{equation*}
which yields
\begin{align*}
\varphi(x_{j_\ast}(t_k+ \frac{\delta}{2 L_\ast}))  \leq d_\star + \epsilon + \left[2(N-1)a^\ast \epsilon -\frac{a_\ast m \delta}{4c_2}\right]\frac{\delta}{2 L_\ast}.
\end{align*}
It is then straightforward to see that
\begin{equation*}
\varphi(x_{j_\ast}(t_k+ \frac{\delta}{2 L_\ast})) < d_\star-\frac{a_\ast m \delta^2}{32 c_2 L_\ast}
\end{equation*}
if we take
\[
\epsilon < \min \Big\{ \frac{a_\ast m \delta^2}{32 c_2 L_\ast}, \frac{a_\ast m \delta}{16(N-1) a^* c_2} \Big\}.
\]
However, this contradicts  the definition of $\varphi$-synchronization since $t_k$ is arbitrarily chosen. This completes the proof and the desired conclusion holds.

\subsection{Proof of Theorem~\ref{unstable-thm1}}\label{proof-unstable-thm1}
The proof is based on the convergence analysis of the scalar quantity
\begin{equation}\label{unstable-lyap-edge}
V(t,x(t))=\max_{\{i,j\}\in\mathcal{V}\times\mathcal{V}}V_{ij}(t,x(t)),
\end{equation}
where
\begin{equation}\label{unstable-lyap-agent-by-agent}
V_{ij}(t,x(t))=\frac{1}{2}e^{-2L(t-t_0)}\|x_i(t)-x_j(t)\|^2, \, \forall \{i,j\}\in\mathcal{V}\times\mathcal{V}.
\end{equation}
Unlike the contradiction argument used for proof of Theorem~\ref{thm1-leaderless}, where the convergence rate is unclear, here we explicitly characterize
the convergence rate. The proof relies on the following lemmas.

\begin{lemma}\label{lemma-invariant-unstable}
Let Assumption~\ref{ass-unstable-nonlinear} hold.
Along the multi-agent dynamics \eqref{multi-agent-closed-leaderless}, $V(t,x(t)) $ is non-increasing for all $t \geq 0$.
\end{lemma}

\begin{IEEEproof}
This lemma establishes a critical non-expansive property along the multi-agent dynamics \eqref{multi-agent-closed-leaderless} for the globally Lipschitz case.
The proof follows from the same techniques as those for proving Lemma \ref{lem-invariant-leaderless} by investigating the Dini derivative of $V(t,x(t))$. 

Let $\overline{\mathcal{V}}_1\times\overline{\mathcal{V}}_2$ be the set containing all the node pairs that reach the maximum at time $t$, i.e.,
$\overline{\mathcal{V}}_1(t)\times\overline{\mathcal{V}}_2(t)=\{\{i,j\}\in\mathcal{V}\times\mathcal{V}: V_{ij}(t)=V(t)\}$.
It is not hard to obtain that
\begin{align}
D^+V \nonumber=& \max_{\{i,j\}\in\overline{\mathcal{V}}_1\times\overline{\mathcal{V}}_2}
\left\{e^{-2L(t-t_0)}(x_i-x_j)\T\right.
(f(t,x_i)-f(t,x_j))+e^{-2L(t-t_0)}(x_i-x_j)\T \nonumber \\
& \times \sum_{k_1 \in\mathcal{N}_i(\sigma(t))}a_{ik_1}(t)(x_{k_1}-x_i)-e^{-2L(t-t_0)}(x_i-x_j)\T\sum_{k_2 \in\mathcal{N}_j(\sigma(t))}a_{jk_2}(t)(x_{k_2}-x_j)\nonumber\\
& \hspace{0.4cm} \left.-Le^{-2L(t-t_0)}\|x_i-x_j\|^2\right\}\nonumber\\
\leq & \frac{1}{2}\max_{\{i,j\}\in\overline{\mathcal{V}}_1\times\overline{\mathcal{V}}_2}\left\{e^{-2L(t-t_0)} \sum_{k_1 \in\mathcal{N}_i(\sigma(t))}a_{ik_1}(t)(\|x_j-x_{k_1}\|^2 \right.-\|x_i-x_j\|^2)\nonumber \\
& \hspace{0.4cm} \left. +e^{-2L(t-t_0)} \sum_{k_2 \in\mathcal{N}_j(\sigma(t))}a_{jk_2}(t)   (\|x_i-x_{k_2}\|^2-\|x_i-x_j\|^2)\right\}\nonumber\\
\leq & \max_{\{i,j\}\in\overline{\mathcal{V}}_1\times\overline{\mathcal{V}}_2}
\left\{\sum_{k_1\in\mathcal{N}_i(\sigma(t))}a_{ik_1}(t)(V_{jk_1}-V_{ij})+\sum_{k_2 \in\mathcal{N}_j(\sigma(t))}a_{jk_2}(t)
(V_{ik_2}-V_{ij})\right\} \leq 0,\label{unstable-invariant}
\end{align}
where the first equality follows from Lemma~\ref{lem1} and \eqref{multi-agent-closed-leaderless},
the first inequality follows from  \eqref{ass-unstable-nonlinear} and $ab\leq \frac{a^2+b^2}{2}$ and $-ab\leq \frac{a^2+b^2}{2}$ for all $a, b \in \R^n$, and the second inequality follows from \eqref{unstable-lyap-agent-by-agent}.
\end{IEEEproof}


\begin{lemma}\label{lem-thm-unstable}
Let Assumption~\ref{ass-unstable-nonlinear} hold. Assume that $\mathcal{G}_{\sigma(t)}$ is uniformly jointly strongly connected. Then there exists $0<\tilde{\beta}<1$ such that
\begin{equation*}\label{unstable-lyap-agent}
V_{ij}(\overline{N}T_0, x(\overline{N}T_0))\leq ~\tilde{\beta}V^*, \quad \, \forall \{i,j\}\in\mathcal{V}\times\mathcal{V},
\end{equation*}
where $\overline{N}=N-1$, $T_0$ is given by \eqref{defn-T0} and $V^*=V(t_0,x(t_0))$.
\end{lemma}

\begin{IEEEproof}
The proof is based on the convergence analysis of $V_{ij}(t,x(t)$ for all agent pairs $\{i,j\}\in\mathcal{V}\times\mathcal{V}$
in several steps, which is similar to the proof of Lemma~\ref{lem1-thm1}. Without loss of generality, we assume that $t_0=0$. We also sometimes denote $V(t,x(t))$ and $V_{ij}(t,x(t))$ as $V$ and $V_{ij}$, respectively, for notational simplification.

\noindent Step 1. We begin by considering any agent $i_1\in\mathcal{V}$. 
Since $\mathcal{G}_{\sigma(t)}$ is uniformly jointly strongly connected, we know that
$i_1$ is the root and that there exists a time $t_1$ and an agent $i_2\in\mathcal{V}\setminus\{i_1\}$ such that $(i_1,i_2)\in \mathcal{E}$ during $t\in[t_1,t_1+\tau_D)\subset[0,T_0]$.

We first note that it follows from \eqref{unstable-lyap-edge} and Lemma~\ref{lemma-invariant-unstable} that for all $t\in[0,\overline{N}T_0]$,
\begin{equation}\label{max-bound}
V_{ij}(t,x(t))\leq V(t,x(t)) \leq V^*, ~\forall \{i,j\}\in\mathcal{V}\times\mathcal{V}.
\end{equation}

Taking the derivative of $V_{ij}$ along the trajectories of \eqref{multi-agent-closed-leaderless}, we obtain that for all $t\in[t_1,t_1+\tau_D)$,
\begin{align*}
\dot V_{i_1i_2} =& -Le^{-2L(t-t_0)}\|x_{i_1}-x_{i_2}\|^2+e^{-2L(t-t_0)}(x_{i_1}-x_{i_2})\T \left\{ \sum_{k_1\in\mathcal{N}_{i_1}(\sigma(t))}
a_{i_1k_1}(t)(x_{k_1}-x_{i_1}) \right.
\\ & \left.
-\sum_{k_2\in\mathcal{N}_{i_2}(\sigma(t))}a_{i_2k_2}(t)(x_{k_2}-x_{i_2})+(f(t,x_{i_1})-f(t,x_{i_2}))\right\}
\\ \leq &~\sum_{k_1\in\mathcal{N}_{i_1}(\sigma(t))}\!\!\!\!a_{i_1k}(t)(V_{i_2k_1}-V_{i_1i_2})-a_{i_2i_1}(t)V_{i_1i_2}+\sum_{k_2 \in\mathcal{N}_{i_2}(\sigma(t))\backslash\{i_1\}}a_{i_2k_2}(t)(V_{i_1k_2}-V_{i_1i_2})
\\ \leq &~(N-1)a^*(V^*-V_{i_1i_2})-a_*V_{i_1i_2}+(N-2)a^*(V^*-V_{i_1i_2})
\\ = & ~-\alpha(V_{i_1i_2}-\frac{(2N-3)a^*}{\alpha}V^*),
\end{align*}
where $\alpha=(2N-3)a^*+a_*$. The first inequality follows from 
\eqref{ass-unstable-nonlinear} and 
\eqref{unstable-lyap-agent-by-agent}, while the second inequality follows from \eqref{max-bound}.

It thus follows that
\begin{equation}\label{Vi1i2bound1}
V_{i_1i_2}(t_1+\tau_D,x(t_1+\tau_D))\leq \hat{\alpha}_1V^*,
\end{equation}
where $\hat{\alpha}_1=1-\frac{a_*}{\alpha}(1-e^{-\alpha \tau_D}) \in (0,1)$.

Similarly we obtain that for all $t\in[t_1+\tau_D,\overline{N}T_0]$, $\dot V_{i_1i_2} \leq \overline{\alpha}(V^*-V_{i_1i_2})$, where $\overline{\alpha}=2(N-1)a^*$. It then follows from \eqref{Vi1i2bound1} that
\begin{equation}\label{vi1vi2bound-end}
V_{i_1i_2}(t,x(t)) \leq \alpha_1^*V^*, \quad \, \forall t\in[t_1+\tau_D,\overline{N}T_0],
\end{equation}
where $\alpha_1^*=1-(1-e^{-\alpha \tau_D})\frac{a_*}{\alpha}e^{-\overline{\alpha}\overline{N}T_0} \in (0,1)$.

\vspace*{2mm}

\noindent Step 2.   
Since $\mathcal{G}_{\sigma(t)}$ is uniformly jointly strongly connected, we know that
that there exists a time instant $t_2$ and an arc from $h\in \mathcal{V}_1 \defn \{i_1,i_2\}$ to
$i_3\in\mathcal{V}\setminus \mathcal{V}_1$ during $[t_2,t_2+\tau_D)\subset[T_0,2T_0]$.

We then estimate an upper bound for $V_{hi_3}$ by considering two different cases: $h=i_1$ and $h=i_2$.
We eventually obtain that for all $t\in[t_2+\tau_D,\overline{N}T_0]$,
\begin{equation}\label{V1V3bound-overall}
V_{h i_3}(t,x(t))\leq ~(1-\beta_*^2)V^*, \quad \forall h \in \mathcal{V}_1.
\end{equation}
where
\begin{equation}\label{beta-star}
\beta_*=(1-e^{-\alpha \tau_D})\frac{a_*}{\alpha}e^{-\overline{\alpha}\overline{N}T_0} \in (0,1).
\end{equation}
It then follows from \eqref{vi1vi2bound-end}, \eqref{V1V3bound-overall}
and $1-\beta_*^2>1-\beta_*=\alpha_1^*$ that for all $t\in[2T_0,\overline{N}T_0]$,
\begin{equation*}
V_{i_1k}(t, x(t)) \leq (1-\beta_*^2)V^*, \quad \forall k\in \mathcal{V}_2\backslash \{i_1\},
\end{equation*}
where $\mathcal{V}_2 \defn \{i_1,i_2,i_3\}$.

\vspace*{2mm}

\noindent Step 3. By continuing the above process, we obtain that for all $k\in \mathcal{V}\backslash \{i_1\}$,
\begin{equation}\label{eq:bound1}
V_{i_1k}(\overline{N}T_0, x(\overline{N}T_0))\leq ~(1-\beta_*^{\overline{N}})V^*.
\end{equation}

\vspace*{2mm}

\noindent Step 4. Since $\mathcal{G}_{\sigma(t)}$ is uniformly jointly strongly connected, \eqref{eq:bound1} holds for any $i_1\in \mathcal{V}$. By using the same analysis, we eventually obtain that for all $i,j\in \mathcal{V}$,
$$
V_{ij}(\overline{N}T_0, x(\overline{N}T_0))\leq ~(1-\beta_*^{\overline{N}})V^*.$$
Hence the result follows by choosing $\tilde{\beta}=1-\beta_*^{\overline{N}}$ with $\beta_*$ given by \eqref{beta-star}.
\end{IEEEproof}

We are now ready to prove Theorem~\ref{unstable-thm1}.
By using Lemma~\ref{lem-thm-unstable} and \eqref{unstable-lyap-edge}, we obtain that
\begin{equation*}
V(t, x(t)) \leq \tilde{\beta}^{\lfloor{{\frac{t}{\overline{N}T_0}}}\rfloor}V^*
\leq \frac{1}{\tilde{\beta}} e^{-\rho_* t}V^*,
\end{equation*}
where $\lfloor{\frac{t}{\overline{N}T_0}}\rfloor$ denotes the largest integer that is not greater than
$\frac{t}{\overline{N}T_0}$ and $\rho_* =\frac{1}{\overline{N}T_0}\ln\frac{1}{\tilde{\beta}}$.

It then follows from \eqref{unstable-lyap-edge} and \eqref{unstable-lyap-agent-by-agent} that
\begin{align*}
\max_{\{i,j\}\in\mathcal{V}\times\mathcal{V}}\|x_i(t)-x_j(t)\|^2\leq \frac{1}{\tilde{\beta}}e^{-(\rho_*-2L)t} \max_{\{i,j\}\in\mathcal{V}\times\mathcal{V}}\|x_i(0)-x_j(0)\|^2.
\end{align*}
Hence, global exponential synchronization is achieved with $\gamma=\frac{1}{\tilde{\beta}}$
and $\lambda=\rho_*-2L$ provided that $\rho_*>2L$.
This concludes the proof of the desired theorem.

\section{Leader-follower Networks}\label{sec-lf}

Our focus so far has been on achieving synchronization for leaderless networks.
In this section we consider the synchronization problem for leader-follower networks.
Suppose that there is an additional agent, labeled as agent $0$,
which plays as a reference or leader for the agents in the set $\mathcal{V}$.
In view of this we also call an agent in $\mathcal{V}$ a follower, and denote  $\bar{\mathcal{V}}=\mathcal{V} \cup \{0\}$ as the overall agent set. The overall communication in the network is described by a time-varying directed graph $\bar{\mathcal{G}}_{\sigma(t)}=(\bar{\mathcal{V}}, \bar{\mathcal{E}}_{\sigma(t)})$. Here for the sake of simplicity we continue to use $\sigma(\cdot)$ to denote the piecewise constant graph signal.
We also make a standard dwell time assumption \cite{liberzon-morse} on  the switching signal $\sigma(t)$. 

For the leader-follower communication graph, we introduce the following definition.
\begin{definition}\label{def-connected-leader}
(i). $\bar{\mathcal{G}}_{\sigma(t)}$ is \emph{leader connected} if
for any follower agent $i \in \mathcal{V}$
there is a directed path from the leader $0$ to follower agent $i$ in $\bar{\mathcal{G}}_{\sigma(t)}$
at time $t$.
Moreover, $\bar{\mathcal{G}}_{\sigma(t)}$ is \emph{jointly leader connected} in the time interval $[t_1,t_2)$ if the union graph $\bar{\mathcal{G}}([t_1,t_2))$ is leader connected.

(ii). $\bar{\mathcal{G}}_{\sigma(t)}$ is {\emph{uniformly jointly leader connected}} if there exists a constant $T>0$ such that the union graph $\bar{\mathcal{G}}([t,t+T))$ is leader connected for any $t \geq 0$.

(iii). $\bar{\mathcal{G}}_{\sigma(t)}$ is {\emph{infinitely jointly leader connected}} if the union graph $\bar{\mathcal{G}}([t, \infty))$ is leader connected for any $t \geq 0$.
\end{definition}

For leader-follower  networks, the evolutions of the follower state $x_i(t)$ and the leader state
$y(t)$ are given by
\begin{equation}\label{multi-agent-closed}
\begin{cases}
\dot{x}_i= f(t,x_i) + \sum _{j \in \mathcal{N}_i(\sigma(t))} a_{ij}(t)(x_j-x_i)+b_i(t)(y-x_i), \quad i \in \mathcal{V}, \\
\dot{y} = f(y,t),
\end{cases}
\end{equation}
where the nonlinear function $f(x_i,t)$ and $a_{ij}(t)$ follow from the same definitions as those of the leaderless case in \eqref{multi-agent-closed-leaderless}, and $b_i(t)>0$ is a piecewise continuous function marking  the strength of the edge $(0,i)$, if any. Assume that there is a constant $b_*>0$ such that $b_* \leq b_i(t)$ for all $t \geq 0$. We also 
assume that the initial time is $t=t_0\geq 0$ and denote the initial state for the leader as $y(t_0) \in {\R}^n$.

For the leader-follower networks, we are interested in the following synchronization problems.

\begin{definition}\label{lf-syn-asy}
(i) The multi-agent system \eqref{multi-agent-closed} achieves global asymptotic synchronization if
$\lim_{t\to \infty}(x_i(t)-y(t))=0$ for any $i \in\mathcal{V}$, any $t_0 \geq 0$, any $x(t_0) \in {\R}^{nN}$, and $y(t_0) \in {\R}^n$.

(ii) Multi-agent system \eqref{multi-agent-closed} achieves global exponential synchronization if there exist $\gamma \geq 1$ and $\lambda>0$ such that
there exist $\gamma \geq 1$ and $\lambda>0$ such that
\begin{equation}\label{global-expo-LF}
\max_{i \in \mathcal{V}} \|x_i(t)-y(t)\|^2 \leq \gamma e^{-\lambda(t-t_0)} \max_{i \in \mathcal{V}}\|x_i(t_0)-y(t_0)\|^2, \quad t\geq t_0,
\end{equation}
for any $t_0 \geq 0$, any $x(t_0) \in {\R}^{nN}$, and $y(t_0) \in {\R}^n$.
\end{definition}

\subsection{Non-expansive Inherent Dynamics}
In this section, we extend the results for the case when the agent dynamics is non-expansive to leader-follower networks. We make the following assumption on the agent dynamics.
\begin{assumption}\label{ass-lyap-lf}
$\varphi_\ast: {\R}^n \rightarrow \R$ is a continuously differentiable positive definite function
such that the following conditions hold:
\begin{itemize}
\item[(i).] $\lim_{\|\eta\| \rightarrow \infty}\varphi_\ast(\eta)=\infty$;
\item[(ii).] $\langle \nabla \varphi_\ast(\eta-\zeta), f(t,\eta)-f(t,\zeta) \rangle \leq 0$ for all $\eta,\zeta \in {\R}^n$ and $t \geq 0$.
\end{itemize}
\end{assumption}
Assumption~\ref{ass-lyap-lf} is similar to Assumption~\ref{ass-lyap} however with possibly different functions $\varphi_*(\cdot)$  and Assumption~\ref{ass-lyap-lf}(ii) holds in the relative coordinate.

The convexity property guarantees the non-expansive property along the leader-follower multi-agent dynamics \eqref{multi-agent-closed} as shown in the following lemma whose proof is similar to that of Lemma~\ref{lem-invariant-leaderless} and thus omitted.
\begin{lemma}\label{lem-invariant}
Let Assumption~\ref{ass-lyap-lf} hold. Along the leader-follower multi-agent dynamics \eqref{multi-agent-closed},\\ 
$\max_{i \in \mathcal{V}} \varphi_\ast(x_i(t))$ is non-increasing for all $t \geq 0$..
\end{lemma}

We now state our results for the non-expansive case.

\begin{theorem}\label{thm1-leader}
Let Assumption~\ref{ass-lyap-lf} hold.
The multi-agent system \eqref{multi-agent-closed} achieves global asymptotic synchronization if $\bar{\mathcal{G}}_{\sigma(t)}$ is uniformly jointly leader connected.
\end{theorem}

\begin{theorem}\label{thm2-leader}
Let Assumption \ref{ass-lyap-lf} hold.
Assume that $\mathcal{G}_{\sigma(t)}$ is undirected for all $t \geq t_0$.
The multi-agent system \eqref{multi-agent-closed} achieves global asymptotic synchronization is achieved if $\bar{\mathcal{G}}_{\sigma(t)}$ is infinitely jointly leader connected.
\end{theorem}

The proofs of Theorems~\ref{thm1-leader} and~\ref{thm2-leader} are given in Appendices~\ref{app-lf1} and~\ref{app-lf2}, respectively, and based on a generalization of the methods proposed in
\cite{moreau,shi-johansson-siam} however the nonlinear agent dynamics results in a different Lyapunov function $\max_{i \in \mathcal{V}} \varphi_*(\bar{x}_i(t))$. The analysis are based on estimating the scalar function $\varphi_*(\bar{x}_i(t))$ agent by agent and thus yields an estimate of the convergence rate. They are different from the contradiction arguments used in the proofs of Theorems~\ref{thm1-leaderless} and \ref{thm2-leaderless} where the convergence rate is unclear.

\begin{remark}\label{remark-exponential}
If the scalar function $\varphi_*(\cdot)$ satisfies an additional condition, i.e.,
there exist $0<c_1 \leq c_2$, such that
\[
c_1\|\eta\|^2 \leq \varphi_*(\eta) \leq c_2\|\eta\|^2, \, \forall \eta \in {\R}^n,
\]
then Theorem \ref{thm1-leader} leads to global exponential synchronization.
\end{remark}

\begin{remark}
For the leader-follower case, Theorems~\ref{thm1-leader} and \ref{thm2-leader} show that
global asymptotic synchronization is achieved while for the leaderless case, while global asymptotic $\varphi$-synchronization is achieved as shown in Theorems~\ref{thm1-leaderless} and \ref{thm2-leaderless}. For the leader-follower case,
the uniformly jointly leader connected in Theorem \ref{thm1-leader}
requires the leader to be a center node, while for the leaderless case, the uniformly jointly strongly connected
in Theorem \ref{thm1-leaderless} requires every node to be a center node.
Thus, the connectivity condition of the leaderless case is stronger than that of the leader-follower case.
\end{remark}

\subsection{Lipschitz Inherent Dynamics}

In this section, we extend the result for the case when the agent dynamics is globally Lipschitz to leader-follower networks.

Our main result for this case is given in the following theorem whose proof can be found in Appendix~\ref{app-lf3}.

\begin{theorem}\label{unstable-thm2}
Let Assumption~\ref{ass-phi-state-1} hold. Suppose that $\bar{\mathcal{G}}_{\sigma(t)}$ is uniformly jointly leader connected.
The multi-agent system \eqref{multi-agent-closed} achieves global exponential synchronization if $L< \hat{\rho}_*/2$, where $\hat{\rho}_*$ is a constant depending on the
network parameters.
\end{theorem}

\section{Conclusions}\label{sec-conclusion}
In this paper, synchronization problems for networks with nonlinear inherent agent dynamics and switching topologies
have been investigated. Two types of nonlinear dynamics were considered: non-expansive and globally Lipschitz.
For the non-expansive case, we found that the convexity of the Lyapunov function plays a crucial rule in the analysis
and showed that the uniformly joint strong connectivity is sufficient for achieving global asymptotic $\varphi$-synchronization.
When communication graphs are undirected, the infinitely joint connectivity is a sufficient synchronization condition.
Moreover, we established conditions under which $\varphi$-synchronization implies state synchronization.
For the globally Lipschitz case, we found that joint connectivity alone is not sufficient to achieve synchronization
but established a sufficient synchronization condition. The proposed condition reveals the relationship between the Lipschitz constant and the network parameters. 
The results were also extended to leader-follower networks.
An interesting future direction is to study the synchronization problem for coupled non-identical nonlinear inherent dynamics under general switching topologies.

\bibliography{IEEEabrv,referenc}
\bibliographystyle{IEEEtran}

\appendix

\section{Proof of Theorem~\ref{thm1-leader}}\label{app-lf1}
The proof of Theorems~\ref{thm1-leader} relies on the following lemma whose proof is similar to that of Lemmas~\ref{lem1-thm1}.
\begin{lemma}\label{lem1-thm3}
Let Assumption~\ref{ass-lyap-lf} hold.  Assume that $\bar{\mathcal{G}}_{\sigma(t)}$ is uniformly jointly leader connected.
Then there exists $0<\bar{\rho}_*<1$ such that
\[
\varphi_*(\bar{x}_i(t_0+T^*)) \leq \bar{\rho}_* \max_{i \in \mathcal{V}}\varphi_*(\bar{x}_i(t_0)), \quad \, \forall i \in \mathcal{V},
\]
where $T^*=NT_0$, with $T_0$ given in \eqref{defn-T0}.
\end{lemma}

\begin{IEEEproof}
Without loss of generality, we assume the initial time $t_0=0$.
Similar to the proof of Lemma~\ref{lem1-thm1}, we estimate $\varphi_*(\bar{x}_i(t))$ agent by agent on the subintervals $t \in [(j-1)T_0, jT_0]$ for $j=1,\ldots,N$ in several steps.

\noindent Step 1. In this step, we focus on an follower agent $k_1 \in \mathcal{V}$ such that
$(0,k_1) \in \bar{\mathcal{E}}_{\sigma(t)}$ for $t \in [t_1,t_1+\tau_D) \subseteq [0,T_0]$.
The existence of such a follower agent $k_1$ and $t_1$ and follows from the fact
$\bar{\mathcal{G}}_{\sigma(t)}$ is uniformly jointly leader connected. 
For convenience, we introduce $\bar{x}_i=x_i-y$ for all $i \in \mathcal{V}$ as the relative state from the leader agent.
From \eqref{multi-agent-closed}, we obtain the following dynamics.
\begin{equation}\label{dynamics-relative}
\dot{\bar{x}}_i=f(t,x_i)-f(t,y)+ \sum _{j \in \mathcal{N}_i(\sigma(t))} a_{ij}(t)(\bar{x}_j-\bar{x}_i)-b_{i}(t)\bar{x}_i.
\end{equation}

Similar to \eqref{leaderless-k1}, we obtain that for all $t \in [t_1, t_1+\tau_D]$,
\begin{align*}
\frac{d}{dt} \varphi_*(\bar{x}_{k_1}(t))\leq a^* (N-1)\left(\max_{i \in \mathcal{V}}\varphi_*(\bar{x}_i(0))-\varphi_*(\bar{x}_{k_1})\right)-b_* \varphi_*(\bar{x}_{k_1}).
\end{align*}
We then obtain that 
\begin{equation}\label{dis-k1}
\varphi_*(\bar{x}_{k_1}(t_1+\tau_D)) \leq \hat{\delta}_1 \max_{i \in \mathcal{V}}\varphi_*(\bar{x}_{k_1}(0)),
\end{equation}
where
\begin{equation}\label{hatdelta1}
\hat{\delta}_1=\frac{\hat{\lambda}_1-b_*(1- e^{-\hat{\lambda}_1\tau_D})}{\hat{\lambda}_1},
\end{equation}
with
\begin{equation}\label{def-lambda1hat}
\hat{\lambda}_1=a^*(N-1)+b_*.
\end{equation}
For $t \in [t_1+\tau_D, T^*]$, the edge $(0,k_1)$ may no longer exist.
Nevertheless, we have for $t \in [0, T^*]$,
\begin{equation}
\frac{d}{dt}  \varphi_*(\bar{x}_{k_1}(t)) \leq a^* (N-1)\left(\max_{i \in \mathcal{V}}\varphi_*(\bar{x}_i(0))-\varphi_*(\bar{x}_{k_1})\right). \label{same-worst}
\end{equation}
It then follows from \eqref{dis-k1} and \eqref{same-worst} that
for all $t \in [T_0, T^*]$,
\begin{equation}\label{vk1}
\varphi_*(\bar{x}_{k_1}(t))  \leq \delta_1 \max_{i \in \mathcal{V}}\varphi_*(\bar{x}_{k_1}(0)),
\end{equation}
where $\delta_1 = 1-e^{-\bar{\lambda}_1 N T_0}(1-\hat{\delta}_1)$ and
$\bar{\lambda}_1=a^* (N-1)$.

\vspace*{2mm}

\noindent Step 2. In this step, we analysis a follower agent $k_2 \in \mathcal{V} \setminus \{k_1\}$ such that
there either an edge $(0,k_2)$ or an edge $(k_1,k_2)$ in $\bar{\mathcal{E}}_{\sigma(t)}$ for $t \in [t_2,t_2+\tau_D) \subseteq [T_0, 2T_0]$. Again, the existence of $k_2$ and $t_2$ due to the uniform joint leader connectivity.
By going through the similar analysis as Step 2 of the proof for Lemma~\ref{lem1-thm1}, we eventually obtain
that for $t \in [2T_0, T^*]$,
\begin{equation*}
\varphi_*(\bar{x}_{j}(t)) < \delta_2 \max_{i \in \mathcal{V}}\varphi_*(\bar{x}_{i}(0)), \quad j \in \{k_1, k_2\},
\end{equation*}
where $\delta_2=1-e^{-\bar{\lambda}_1 (N-1)T_0}(1-\hat{\delta}_2)$,
with
\begin{equation*}\label{hatdelta2}
\hat{\delta}_2= \frac{\bar{\lambda}_2-a_*(1-\delta_1) (1-e^{-\bar{\lambda}_2 \tau_D})}{\bar{\lambda}_2}
\end{equation*}
and $\bar{\lambda}_2=a^* (N-2)+a_*$.

\vspace*{2mm}

\noindent Step 3. By applying the similar analysis on the subintervals $[(\ell-1)T_0,\ell T_0]$ for $\ell=3, \ldots, N$,
we obtain that for all $t \in [\ell T_0, T^*]$ and for all $j \in \{k_1, \ldots, k_\ell \}$,
\begin{equation} \label{allless}
\varphi_*(\bar{x}_{j}(t))  < \delta_\ell \max_{i \in \mathcal{V}}\varphi_*(\bar{x}_i(0)),
\end{equation}
where
\begin{equation}\label{deltai}
\delta_\ell=1-e^{-\bar{\lambda}_1(N-\ell+1)T_0}(1-\hat{\delta}_\ell),
\end{equation}
and
\begin{equation}\label{hatdeltai}
\hat{\delta}_\ell= \frac{a^*(N-2)+a_*-a_*(1-\delta_{\ell-1}) (1-e^{-\bar{\lambda}_2 \tau_D})}{a^*(N-2)+a_*}.
\end{equation}
It follows from \eqref{deltai} and \eqref{hatdeltai} that $\delta_{k} \leq \delta_{N}$ for all $k=2,\ldots,N$.
This together with \eqref{allless} leads to
$\varphi_*(\bar{x}_{j}(T^*)) \leq \delta_N \max_{i \in \mathcal{V}}\varphi_*(\bar{x}_i(0))$ for all $j \in \mathcal{V}$, where
\begin{equation}\label{deltaN}
\delta_N=1-\eta^{N-1}e^{\frac{(N-2)(N+1)}{2}\bar{\lambda}_1 T_0}b_*(1-e^{-\hat{\lambda}_1 \tau_D})/({a^*(N-1)+b_*}),
\end{equation}
with
\[
\eta=\frac{e^{-\bar{\lambda}_1(N+1)T_0}(1-e^{-\bar{\lambda}_2\tau_D})a_*  }{a^*(N-2)+a_*}.
\]
The result of lemma then follows by choosing $\bar{\rho}_*=\delta_{N}$.
\end{IEEEproof}

We are ready to prove Theorem~\ref{thm1-leader}. Without loss of generality, we assume that $t_0=0$, it then follows from Lemma~\ref{lem1-thm3} that $\max_{i \in \mathcal{V}}\varphi_*(\bar{x}_i(T^*)) \leq \bar{\rho}_* \max_{i \in \mathcal{V}}\varphi_*(\bar{x}_i(0))$. Thus, for $s=0,1,\ldots$, we have
\[
\max_{i \in \mathcal{V}}\varphi_*(\bar{x}_i(sT^*)) \leq \bar{\rho}_* ^s \max_{i \in \mathcal{V}}\varphi_*(\bar{x}_i(0)).
\]
It then follows that
\begin{equation*}
\max_{i \in \mathcal{V}}\varphi_*(\bar{x}_i(t)) \leq {\bar{\rho}}^{\frac{t}{T^*}-1} _* \max_{i \in \mathcal{V}}\varphi_*(\bar{x}_i(0)),
\end{equation*}
This together with the fact that $\varphi_*(\cdot)$ is positive definite as given in Assumption~\ref{ass-lyap-lf} implies that
the multi-agent system \eqref{multi-agent-closed} achieves global asymptotic synchronization.

\section{Proof of Theorem~\ref{thm2-leader}}\label{app-lf2}
The proof of Theorems~\ref{thm2-leader} relies on the following lemma whose proof is similar to that of Lemmas~\ref{lem1-thm1}.

\begin{lemma}\label{lem1-thm4}
Let Assumption~\ref{ass-lyap-lf} hold. Assume that $\bar{\mathcal{G}}_{\sigma(t)}$ is infinitely jointly leader connected.
Then there exist $0<\tilde{\rho}_*<1$, $T_p$ and $\tilde{t}_*$ such that
\[
\varphi_*(\bar{x}_i(\tilde{t}_*+\tau_D)) \leq \tilde{\rho}_* \max_{i \in \mathcal{V}}\varphi_*(\bar{x}_i(T_p)), \quad \, \forall i \in \mathcal{V}.
\]
\end{lemma}

\begin{IEEEproof}
Since $\bar{\mathcal{G}}_{\sigma(t)}$ is infinitely jointly leader connected, there exist a sequence of time instants
\begin{equation}\label{defnTp1}
0=T_0<T_1< \ldots < T_p < T_{p+1}< \ldots
\end{equation}
such that
\begin{equation}\label{defnTp2}
T_p \defn t_{p_1}< t_{p_2} <\ldots< t_{p_{N+1}} \defn T_{p+1}
\end{equation}
for $p=0,1,\ldots$, and $\bar{\mathcal{G}}([t_{p_\ell},t_{p_{\ell+1}}))$ is leader connected for $\ell=1,\ldots, N$.
Moreover, each edge in $\bar{\mathcal{G}}([t_{p_\ell},t_{p_{\ell+1}}))$
exists for at least the dwell time $\tau_D$ during $[t_{p_\ell},t_{p_{\ell+1}})$
for $p=0,1,\ldots$ and $\ell=1,\ldots, N$.

We shall estimate $\varphi_*(\bar{x}_i(t))$ agent by agent on the subintervals $[t_{p_\ell},t_{p_{\ell+1}}]$, $\ell=1,\ldots, N$
for the interval $[T_p,T_{p+1}]$, $p=0,1,\ldots$.

\noindent Step 1. In this step, we focus all the follower agent $i \in \mathcal{V}_1$, where
\[
\mathcal{V}_1 \defn \{ i \in \mathcal{V} \mid (0,i) \in \bar{\mathcal{E}}_{\sigma(t_1)}\}.
\]
and
\[
t_1 \defn \inf_{t \in [t_{p_1},t_{p_2})} \{\exists \, i \in \mathcal{V} \mid (0,i) \in \bar{\mathcal{E}}_{\sigma(t)}\}.
\]
The existence of $\mathcal{V}_1$ and $t_1$ due to the fact that $\bar{\mathcal{G}}([t_{p_1},t_{p_2}))$ is leader connected.
It then follows from the similar analysis as we obtained \eqref{dis-k1} that
\begin{equation}\label{phibark1}
\varphi_*(\bar{x}_{k_1}(t_1+\tau_D)) \leq \hat{\delta}_1 \max_{i \in \mathcal{V}}\varphi_*(\bar{x}_i(T_p)), \quad \forall k_1 \in \mathcal{V}_1,
\end{equation}
where $\hat{\delta}_1$ is given by \eqref{hatdelta1}.

\vspace*{2mm}

\noindent Step 2. In this step, similar to Step 3 of the proof for Lemma~\ref{lem1-thm2},
we view the set $\{0\} \cup \mathcal{V}_1$ as a subsystem. Define $t_2$ as the first time when
there is an edge between this subsystem and the remaining follower agents and $\mathcal{V}_2$ accordingly.

By going through the similar analysis as Step 2 of the proof for Lemma~\ref{lem1-thm1},
we eventually obtain that for $j \in \mathcal{V}_1 \cup \mathcal{V}_2$,
\[
\varphi_*(\bar{x}_{j}(t_2+\tau_D)) < \tilde{\delta}_2 \max_{i \in \mathcal{V}}\varphi_*(\bar{x}_{j}(T_p)),
\]
where
\begin{equation}\label{tildedelta2}
\tilde{{\delta}}_2 =\frac{\bar{\lambda}_2-a_* e^{-\bar{\lambda}_1 \tau_D}(1-\tilde{\delta}_1) (1-e^{-\bar{\lambda}_2 \tau_D})}{\bar{\lambda}_2},
\end{equation}
with
\begin{equation}\label{tildedelta1}
\tilde{\delta}_1=1-e^{-\bar{\lambda}_1\tau_D}(1-{\hat{\delta}}_1),
\end{equation}
and $\hat{\delta}_1$ given by \eqref{hatdelta1}.

\vspace*{2mm}

\noindent Step 3. Since $\bar{\mathcal{G}}_{\sigma(t)}$ is infinitely jointly leader connected,
we proceed the above analysis until $\mathcal{V}=\mathcal{V}_1 \cup \ldots \cup \mathcal{V}_{m_0}$
for some $m_0 \leq N$ such that
\begin{equation*}
\varphi_*(\bar{x}_{j}(t_{m_0}+\tau_D)) < \tilde{\delta}_{m_0} \max_{i \in \mathcal{V}}\varphi_*(\bar{x}_i(T_p)), \quad \forall j \in \mathcal{V},
\end{equation*}
with $t_{m_0}$ defined similarly to $t_1$ and $t_2$, and
\begin{equation*}\label{tildedeltaell}
\tilde{{\delta}}_\ell =\frac{\bar{\lambda}_2-a_*e^{-\bar{\lambda}_1 \tau_D}(1-\tilde{\delta}_{\ell-1}) (1-e^{\bar{\lambda}_2 \tau_D})}{\bar{\lambda}_2}, \quad \ell=3, \ldots, m_0.
\end{equation*}
From the preceding relation and \eqref{tildedelta2},
we obtain for $\ell=2,\ldots, m_0$,
\[
\frac{1-\tilde{\delta}_{\ell}}{1-\tilde{\delta}_{\ell-1}}=\frac{a_*e^{-\bar{\lambda}_1 \tau_D} (1-e^{-\bar{\lambda}_2 \tau_D})}{a^*(N-2)+a_*}\defn \tilde{\eta}<1.
\]
It is then easy to see that $\tilde{\delta}_{\ell-1} < \tilde{\delta}_{\ell}$ for all $\ell=2, \ldots, m_0$.
This together with \eqref{hatdelta1}, \eqref{tildedelta1}, and $m_0 \leq N$ implies that for all  $j \in \mathcal{V}$,
\[
\varphi_*(\bar{x}_j(t_{m_0}+\tau_D)) \leq \tilde{\delta}_{N} \max_{i \in \mathcal{V}}\varphi_*(\bar{x}_i(T_p)),
\]
where
\[
\tilde{\delta}_N =1- \tilde{\eta}^{N-1}b_*e^{-\bar{\lambda}_1 \tau_D}(1-e^{-\hat{\lambda}_1\tau_D})/({a^*(N-1)+b_*}),
\]
with $\hat{\lambda}_1$ given by \eqref{def-lambda1hat}.
The result then follows by choosing $T_p$ as defined in \eqref{defnTp1} and \eqref{defnTp2}, $\tilde{t}_*=t_{m_0}$ and $\tilde{\rho}_*= \tilde{\delta}_N$.
\end{IEEEproof}

We are ready to prove Theorem~\ref{thm1-leader}. Without loss of generality, we assume that $t_0=0$, it follows then follows from Lemma~\ref{lem-invariant}, Lemma~\ref{lem1-thm4}, and the fact that
$\tilde{t}_*+\tau_D \leq T_{p+1}$ for $\tilde{t}_*=t_{m_0}$, which follows from the definition of $T_{p+1}$ given in \eqref{defnTp1} and \eqref{defnTp2}, that for $p=0,1,\ldots$
\[
\max_{i \in \mathcal{V}}\varphi_*(\bar{x}_i(T_{p+1})) \leq  \max_{i \in \mathcal{V}}\varphi_*(\bar{x}_i(t_{m_0}+\tau_D)) \leq \tilde{\rho}_*  \max_{i \in \mathcal{V}}\varphi_*(\bar{x}_i(T_p)).
\]
Thus, we have for $s=0,1,\ldots$, $\max_{i \in \mathcal{V}}\varphi_*(\bar{x}_i(T_{s})) \leq \tilde{\rho}^s_*   \max_{i \in \mathcal{V}}\varphi_*(\bar{x}_i(0))$. This together with the fact that $\varphi_*(\cdot)$ is positive definite as given in Assumption~\ref{ass-lyap-lf} implies that the multi-agent system \eqref{multi-agent-closed} achieves global asymptotic synchronization.

\section{Proof of Theorem~\ref{unstable-thm2}}\label{app-lf3}
The proof is similar to that of Theorem~\ref{unstable-thm1} however in the relative coordinate $\bar{x}_i=x_i-y$ whose evolution is given by \eqref{dynamics-relative}, Again, without loss of generality, we assume the initial time $t_0=0$.

The proof is based on the convergence analysis of the nonnegative scalar
\begin{equation}\label{unstable-lyap}
V(t,\bar{x}(t))=\max_{i\in\mathcal{V}}V_i(t,\bar{x}_i(t)),
\end{equation}
where $\bar{x}(t)=[\bar{x}_1\T(t),\bar{x}_2\T(t),\dots,\bar{x}_N\T(t)]\T$ and
\begin{equation}\label{unstable-agent-lyap}
V_i(t,\bar{x}_i(t))=\frac{1}{2}e^{-2L t} \|\bar{x}_i(t)\|^2, \quad \forall i\in\mathcal{V}.
\end{equation}

Let us define $\mathcal{I}(t)=\{i\in\mathcal{V}: V_{i}(t,\bar{x}_i(t))=V(t,\bar{x}(t))\}$.
Similar to Lemma~\ref{lemma-invariant-unstable}, we obtain that $D^+V(t,\bar{x}(t)) \leq 0$ for all $t \geq 0$ along the multi-agent dynamics~\eqref{multi-agent-closed}.
By combining the preceding relation 
we have $V_i(t,\bar{x}_i)\leq V(t,\bar{x}) \leq V(t_0,\bar{x}(0)) \defn V_*$, for all $t\geq t_0$ and all $i\in\mathcal{V}$.
Following the similar analysis as the proof of Theorem \ref{thm1-leader}, we can show that
\begin{equation*}
V(t,\bar{x}(t))\leq \frac{1}{\delta_N} e^{-\hat{\rho}_* t}V_*,
\end{equation*}
where $\delta_N$ is given by  \eqref{deltaN} and $\hat{\rho}_*$ is given by \eqref{GES-paras},
\begin{equation}\label{GES-paras}
\hat{\rho}_*=\frac{1}{T^{*}} \ln \frac{1}{\delta_N}.
\end{equation}
with $T^*=(N-1)T_0$. It then follows that
\begin{align*}
\max_{i\in\mathcal{V}}\|\bar{x}_i(t)\|^2\leq &~\frac{1}{\delta_N} e^{-(\hat{\rho}_*-2L)t}\max_{i\in\mathcal{V}}\|\bar{x}_i(0)\|^2.
\end{align*}
Hence, global exponential synchronization is achieved with $\gamma=\frac{1}{\delta_N}$ and $\lambda=\hat{\rho}_*-2L$ provided that $\hat{\rho}_*>2L$. 

\end{document}